\documentclass[aps,amsmath,amssymb,twocolumn]{revtex4}
\usepackage{graphicx}

\newcommand{\p}{\partial}
\newcommand{\eps}{\epsilon}
\newcommand{\ve}{\varepsilon}
\newcommand{\cE}{{\mathcal E}}

\newcommand{\tr}{\textrm{tr}}

\begin{document}

\title{Glueballs and the universal energy spectrum of tight knots and
links}\thanks{Talk given by TWK at Coral Gables 2003 in celebration of
Paul Frampton's 60th birthday}

\author{Roman V. Buniy}
\email{roman.buniy@vanderbilt.edu}
\affiliation{Department of Physics and Astronomy, Vanderbilt
University, Nashville, TN 37235, USA}

\author{Thomas W. Kephart} 
\email{thomas.w.kephart@vanderbilt.edu}
\affiliation{Department of Physics and Astronomy, Vanderbilt
University, Nashville, TN 37235, USA}

\begin{abstract}
Systems of tightly knotted, linked, or braided flux tubes will have a
universal mass-energy spectrum if the flux is quantized.  We focus on
a model of glueballs as knotted QCD flux tubes.
\end{abstract}

\maketitle

Plasma physics informs us that linked magnetic flux tubes are much
more stable than an unknotted single
loop~\cite{Moffatt:1985}. Linked and knotted flux tubes carry
topological quantum numbers, and one can think of a knot as a
self-linked loop.  Similar comments apply to braids.

Our interest will be in tubes carrying quantized flux in tight knot
and link configurations.  If the loops (tubes) have fixed uniform
thickness and circular cross-section, then each knot and link has a
completely specified length if the configuration is tight, i.e., is of
the shortest length with the tubes non-overlapping and their
cross-sections undistorted. If tubes have uniform cross sections, as
can be approximately the case for many physical systems, then the
length of the tight knot is proportional to the mass (or energy) of
the knot. This, we claim, generates a universal mass (energy) spectrum
for knotted/linked configurations of objects of this type. The lengths
of tight knots were not studied until the mid-1990s~\cite{bio}, and
only recently have accurate calculations of large numbers of tight
knots~\cite{Rawdon} and links~\cite{BKPR} become available. These
results now make it possible to examine physical systems and compare
them with the knot spectrum. We have examined the glueball spectrum of
QCD~\cite{Buniy:2002yx}, \cite{SFLD}. Glueballs~\cite{PDG} are likely
to be solitonic states (See Ref.~\cite{Buniy:2002yx} for detailed
references.) that are solutions to the QCD field equations. While QCD
will be our main focus, there are many more systems where tight knots
may play a role.

In order to decide if a system of flux tubes falls into the universal
class of having a tight knot energy spectrum, we must first
investigate the time scales involved. These are the lifetime of the
soliton $\tau_s$ and the relaxation time $\tau_r$ necessary to reach
the ground state of a knotted configuration (i.e., the tight knot
state). The soliton lifetime (or the corresponding decay width
$\Gamma_s=1/\tau_s$) can depend on several factors. These include the
effects of flux tube breaking, rearrangement, and reconnection. The
partial width for flux tube breaking is non-zero if the production of
particle/anti-particle pairs at the break point is energetically
allowed, for example quark/anti-quark ($q\bar{q}$) pairs for color
electric flux tubes. The partial widths can vary widely depending on
the particle masses.  Rearrangement is a quantum effect where, for
example, in a linked double donut arrangement, the loops can tunnel
free of each other. Finally, reconnection is an effect where tubes
break and re-attach in a different configuration. Such behavior has
been seen in plasma physics and is of major importance in
understanding a variety of astrophysical systems.  All these processes
change topological charge, and their partial widths compete more or
less favorably with each other depending on the parameters that
describe the system.

While no knot lengths have been calculated exactly, it is possible to
calculate the exact lengths of an infinite number of links and
braids~\cite{BKPR}. For links, these calculations are possible in
the case where individual elements of the link lie in planes. For
braids, exact calculations are possible when the elements of the braid
are either straight sections or where their centerlines follow helical
paths.  The shortest of all links, the double donut, is exactly
calculable. The two elements lie in perpendicular planes and are tori
of equal length.  The shortest non--trivial braid is a helically
twisted pair.  ``Weyl's tube formula'' ~\cite{Weyl,Gray} states
that for a tube of constant cross-section $\sigma $ normal to a path
of length $l$, the volume of the tube in flat 3D is just
$V_T=l\sigma$. If we have an analytic form for the path and a circular
cross-section we can find $V_T$. This leads to the class of exactly
calculable links and braids, but since there are no known analytic
forms for the path of tight knots, their volumes can only be
calculated numerically. We can then calculate or estimate the volume
and therefore the energy for a corresponding physical system. As with
knots, the volumes of topologically non-trivial tight braids (those
where the elements are woven together) can only be found
approximately.  While the simple helically twisted braid has a volume
that depends on the pitch angle which can potentially be adjusted by
experimental conditions, tight knots and links have no such adjustable
parameter.
 
Let us begin with a discussion of tight links of flux in
electromagnetic plasma.  Movement of fluids can exhibit topological
properties.  Interrelation between hydro- and magnetic dynamics may
cause magnetic fields, to exhibit topological properties as well. For
example, for a perfectly conducting fluid, the (abelian) magnetic
helicity ${\mathcal H}=\int d ^3 x\,\eps^{ijk}A_i\p_jA_k$ is an
invariant of the motion~\cite{Woltier}, and this quantity can be
interpreted in terms of knottedness of magnetic flux
lines~\cite{Moffatt:1969}.  (The helicity for two linked flux
tubes with fluxes $\Phi_1$ and $\Phi_2$ is ${\mathcal
H}=2n\Phi_1\Phi_2$, where $n$ is the Gauss linking number of the two
tubes.  It is straightforward to generalize this to the case of linked
and/or self-linked thick flux tubes.)  A perfectly conducting,
non-viscous, incompressible fluid relaxes to a state of magnetic
equilibrium without a change in topology~\cite{Moffatt:1985}. The
system approaches a state of magnetic equilibrium by decreasing its
magnetic energy by contraction of the magnetic field lines. In the
case of trivial topology, closed curves contract to a point without
crossing each other.  The relaxation eventually leads to a state with
zero fields (vacuum).  If, however, the topology of the initial
magnetic fields is non-trivial, the relaxation stops when flux tubes
are tightly knotted or linked. This happens because the ``freeze-in''
condition forces topological restrictions on possible changes in field
configurations and so any initial knots and links of field lines
remain topologically unchanged during relaxation. The energy of a
final (equilibrium) state is determined by topology.

By analogy with the abelian case, for a conserved non-abelian
helicity~\cite{Jackiw:2000cd}, we choose the corresponding
expression for helicity with topological properties, ${\mathcal
H}=\int_V \left(A d A+\tfrac{2}{3}A^3\right)$.  In a perfectly
conductive relativistic non-abelian plasma, the electric field
vanishes in the local frame moving with the plasma. Details are
analogous to the abelian case. With these facts in mind, we are now in
position to ask if one could hope to find knotted/linked flux tubes in
a physical system.

For several reasons we believe the ideal physical system in which to
discover and study tight knots and links is Quantum Chromodynamics
(QCD).  These include: (1) QCD is a solidly based part of the standard
model of particle physics, and much about color confinement and the
quark model is already well understood in this context, making much
previous work transferable to the problem of tightly knotted flux
tubes in this theory. (2) Unlike plasmas, fluids or condensed matter
systems where flux tubes are excitations of some media with many
parameters that could hide universal behavior, flux knots in QCD can
exist in the vacuum. Thus continuum states are absent and there are no
media parameters to vary and obscure the universality. Hence, the
results in QCD can be far less ambiguous. (3) The hadronic energy
spectrum has been measured over a large range of energies (140 MeV to
10 GeV) and already many hundreds of states are known. We expect that
among these, a few dozen can be classified as tightly knotted/linked
flux tubes states. These states must have no valance quarks (i.e., no
flavor quantum numbers) in order to be classified as glueballs. (4)
Knotted solitons in QFT are already known to exist. (5) One can
efficiently search for new glueball states at accelerators. (Also,
data from older experiments still exist and can be reanalyzed to check
the predictions of new states described below.)

Consider a hadronic collision that produces some number of baryons and
mesons plus a gluonic state in the form of a closed QCD flux tube (or
a set of tubes). From an initial state, the fields in the flux tubes
quickly relax to an equilibrium configuration, which is topologically
equivalent to the initial state. (We assume topological quantum
numbers are conserved during this rapid process.) The relaxation
proceeds through minimization of the field energy. Flux conservation
and energy minimization force the fields to be homogeneous across the
tube cross sections. This process occurs via shrinking the tube
length, and halts to form a ``tight'' knot or link. The radial scale
will be set by $\Lambda _{\textrm{QCD}}^{-1}$. The energy of the final
state depends only on the topology of the initial state and can be
estimated as follows. An arbitrarily knotted tube of radius $a$ and
length $l$ has the volume $\pi a^2 l$. Using conservation of flux
$\Phi_E$, the energy becomes $\propto l(\tr\Phi_E^2)/(\pi
a^2)$. Fixing the radius of the tube (to be proportional to
$\Lambda_{\textrm{QCD}}^{-1}$), we find that the energy is
proportional to the length $l$. The dimensionless ratio
$\ve(K)=l/(2a)$ is a topological invariant and the simplest definition
of the ``knot energy''~\cite{knot_energy}, and can be used to fit
the correspondence between knot/link energies and glueball masses.

\begin{table*}
\caption{\label{table}Comparison between the glueball mass spectrum
and knot energies.}
\begin{ruledtabular}
\begin{tabular}{ccccc} State & Mass & 
$K$~\footnote{Notation $n^l_k$ means a link of $l$ components with $n$
crossings, and occurring in the standard table of links (see
e.g. \protect\cite{Rolfsen}) on the $k^\textrm{th}$ place. $K\#K'$
stands for the knot product (connected sum) of knots $K$ and $K'$ and
$K*K'$ is the link of the knots $K$ and $K'$.} &
$\ve(K)$~\footnote{Values are from \protect\cite{bio} except for our
exact calculations of $2^2_1$, $2^2_1*0_1$, and $(2^2_1*0_1)*0_1$ in
square brackets, our analytic estimates given in parentheses, and our
rough estimates given in double parentheses.} &
$E(G)$~\footnote{$E(G)$ is obtained from $\ve(K)$ using the fit in
Figure~\ref{figure}.} \\ \hline {\rule[1mm]{0mm}{3mm} $f_0(600)$} &
$400-1200$ & $2^2_1$ & $12.6\ [4\pi]$ & $768\ [766]$\\ $f_0(980)$ &
$980\pm 10$ & $3_1$ & $16.4$ & $993$\\ $f_2(1270)$ & $1275.4\pm 1.2$ &
$2^2_1*0_1$ & $[6\pi+2]$ & $[1256]$\\ $f_1(1285)$ & $1281.9\pm 0.6$ &
$4_1$ & $21.2$ & $1277$\\ & & $4^2_1$ & $(21.4)$ & $(1289)$\\
$f_1(1420)$ & $1426.3\pm 1.1$ & $5_1$ & $24.2$ & $1454$\\$\{f_2(1430)$
& $\approx 1430\}$~\footnote{States in braces are not in the Particle
Data Group (PDG) summary tables.} & $5_1$ & $24.2$ & $1454+\delta'$\\
$f_0(1370)$ & $1200-1500$ & $3_1*0_1$ & $(24.7)$ &
$(1484)$\\$f_0(1500)$ & $1507\pm 5$ & $5_2$ & $24.9$ & $1496$\\
$\{f_1(1510)$ & $1518\pm 5\}$ & $5_2$ & $24.9$ &
$1496+\delta$\\$f'_2(1525)$ & $1525\pm 5$ & $5_2$ & $24.9$ &
$1496+3\delta$\\ $\{f_2(1565)$ & $1546\pm 12\}$ & $5^2_1$ & $(25.9)$ &
$(1555)$\\ $\{f_2(1640)$ & $1638\pm 6\}$ & $6^3_3$ & $((27.3))$ &
$((1638))$\\ \multicolumn{5}{c}{.\dotfill.}\\ & &
$(2^2_1*0_1)*0_1$~\footnote{This is the link product that is not
$2^2_1*2^2_1$.} & $[8\pi+3]$ & $[1686]$~\footnote{Resonances have been
seen in this region, but are unconfirmed~\cite{PDG}.}\\$f_0(1710)$ &
$1713\pm 6$ & $6^3_2$ & $((28.6))$ & $((1714))$\\ & & $3_1\#3_1^*$ &
$28.9\ (30.5)$ & $1732\ (1827)$\\ & & $3_1\#3_1$ & $29.1\ (30.5)$ &
$1744\ (1827)$\\ & & $2^2_1*2^2_1$ & $[8\pi+4]$ & $[1745]$\\ & & $6_2$
& $29.2$ & $1750$\\ & & $6_1$ & $29.3$ & $1756$\\ & & $6_3$ & $30.5$ &
$1827$\\ & & $7_1$ & $30.9$ & $1850$\\ & & $8_{19}$ & $31.0$ &
$1856$\\ & & $8_{20}$ & $32.7$ & $1957$\\ $f_2(2010)$ &
$2011^{+60}_{-80}$ & $7_2$ & $33.2$ & $1986$\\ $f_4(2050)$ & $2025\pm
8$ & $8_{21}$ & $33.9$ & $2028$\\ & & $8_1$ & $37.0$ & $2211$\\ & &
$10_{161,162}$ & $37.6$ & $2247$\\ $f_2(2300)$ & $2297\pm 28$ &
$8_{18}$, $9_1$ & $38.3$ & $2288$\\ $f_2(2340)$ & $2339\pm 60$ & $9_2$
& $40.0$ & $2389$\\ & & $10_1$ & $44.8$ & $2672$\\ & & $11_1$ & $47.0$
& $2802$\\
\end{tabular}
\end{ruledtabular}
\end{table*}

In our model, the chromoelectric fields $F_{0i}$ are confined to
knotted/linked tubes. After an initial time evolution, the system
reaches a static equilibrium state which is described by the energy
density ${\cE}_E=\tfrac{1}{2}\tr F_{0i}F^{0i}-V$. Similar to the bag
model, we have included a constant potential energy $V$ needed to keep
the tubes at a fixed cross-section. The chromoelectric flux $\Phi_E$
is conserved and we assume flux tubes carry one flux quantum. To
account for conservation of the flux, we add the term
$\tr\lambda\{\Phi_E/(\pi a^2)-n^iF_{0i}\}$ to the energy density,
where $n^i$ is the normal vector to a section of the tube and
$\lambda$ is a Lagrange multiplier. The energy density should be
constant under variations of the degrees of freedom, the gauge
potentials $A_\mu$. This leads to a constant field solution,
$F_{0i}=(\Phi_E/\pi a^2)n_i\label{F}$. With this solution, the energy
is positive and proportional to $l$ and thus the minimum of the energy
is achieved by shortening $l$, i.e., tightening the knot.

Lattice calculations, QCD sum rules, electric flux tube models, and
constituent glue models agree that the lightest non--$q\bar{q}$ states
are glueballs with quantum numbers $J^{++}=0^{++}$ and $2^{++}$
~\cite{PDG}.  We will model all $J^{++}$ states (i.e., all
$f_{J}$ and $f'_J$ states listed by the PDG~\cite{PDG}), some of
which will be identified with rotational excitations, as
knotted/linked chromoelectric QCD flux tubes. We proceed to identify
knotted and linked QCD flux tubes with glueballs, where we include all
$f_J$ and $f'_J$ states. The lightest candidate is the $f_0(600)$,
which we identify with the shortest knot/link, i.e., the $2^2_1$ link;
the $f_0(980)$ is identified with the next shortest knot, the $3_{1}$
trefoil knot, and so forth. All knot and link energies have been
calculated for states with energies less then
$1680\,\textrm{MeV}$. Above $1680\,\textrm{MeV}$ the number of knots
and links grows rapidly, and few of their energies have been
calculated ~(see alternatively, Ref.~\cite{BKPR}). However, we do
find knot energies corresponding to all known $f_J$ and $f'_J$ states,
and so can make preliminary identifications in this region.  Our
detailed results are collected in Table \ref{table}, where we list
$f_J$ and $f'_J$ masses, our identifications of these states with
knots and the corresponding knot energies.
\begin{figure*}
\centering \includegraphics[angle=0,width=12cm]{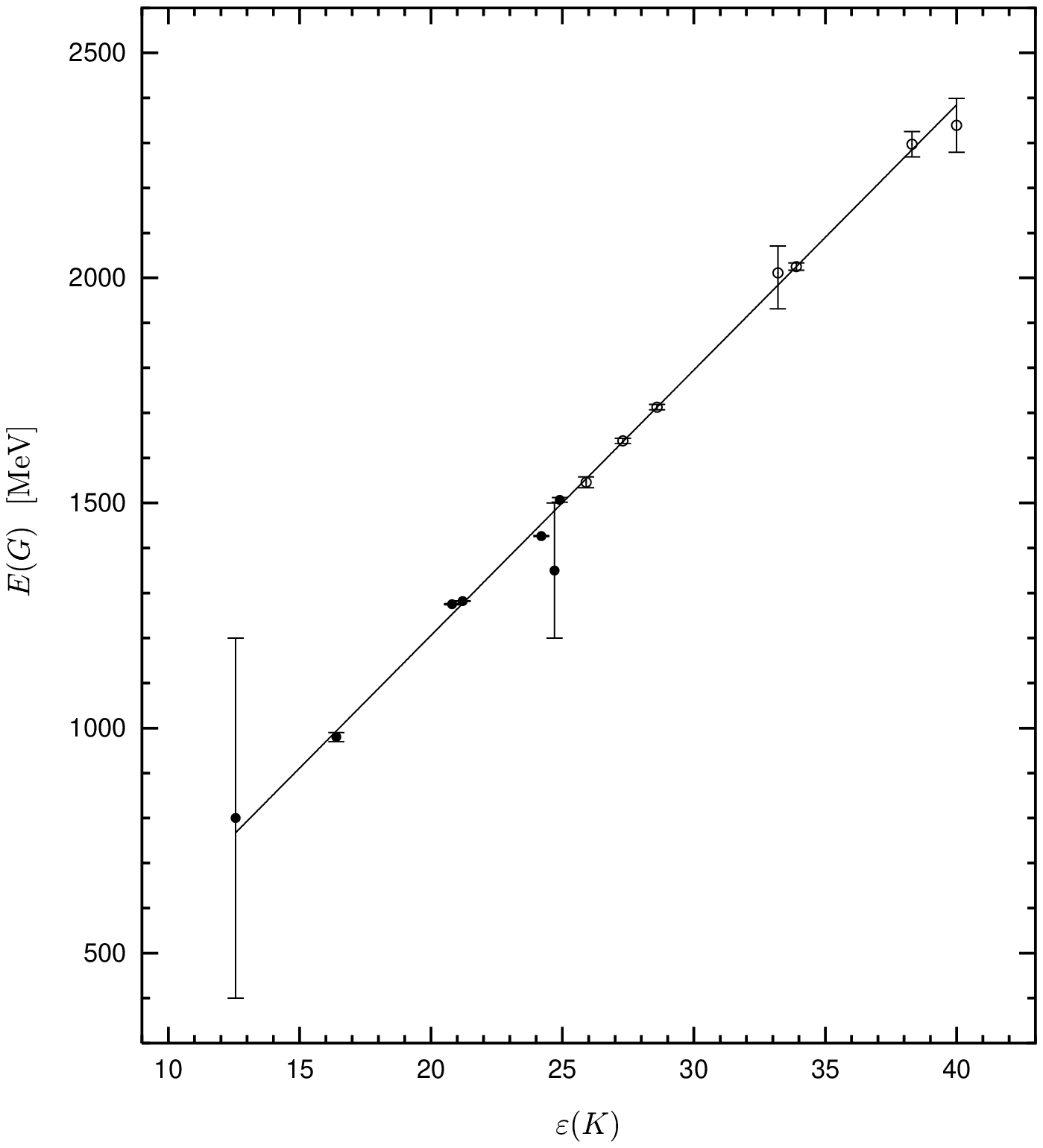}
\caption{\label{figure}Relationship between the glueball spectrum
$E(G)$ and knot energies $\varepsilon(K)$. Each point in this figure
represents a glueball identified with a knot or link. The straight
line is our model and is drawn for the fit $E(G)=(23.4\pm
46.1)+(59.1\pm 2.1)\varepsilon(K)\ \ \ [\textrm{MeV}]$.}
\end{figure*}
In Fig.~\ref{figure} we compare the mass spectrum of $f_J$ states
with the identified knot and link energies. Since errors for the knot
energies in Ref.~\cite{bio} were not reported, we conservatively
assumed the error to be $1\%$. A least squares fit to the most
reliable data (below $1680\,\textrm{MeV}$) gives $E(G)=(23.4\pm
46.1)+(59.1\pm 2.1)\varepsilon(K)\ \ \ [\textrm{MeV}],$ with
$\chi^2=9.1$. The data used in this fit is the first seven $f_J$
states (filled circles in Fig.~\ref{figure}) in the PDG summary
tables. Inclusion of the remaining seven (non-excitation) states
(unfilled circles in Fig.~\ref{figure}) in Table~\ref{table}, where
either the glueball or knot energies are less reliable, does not
significantly alter the fit and leads to $E(G)=(26.9\pm 24.9)+(58.9\pm
1.0)\varepsilon(K)\ \ \ [\textrm{MeV}],$ with $\chi^2=10.1$. Both fits
are in good agreement with our model, where $E(G)$ is proportional to
$\varepsilon(K)$. Better HEP data and the calculation of more knot
energies will provide further tests of the model and improve the high
mass identification.

Knot complexity can be reduced (or increased) by unknotting (knotting)
operations. In terms of flux tubes, these moves are equivalent to
reconnection events. Hence, a metastable glueball may decay via
reconnection. Once all topological charge is lost, metastability is
lost, and the decay proceeds to completion. Two other glueball decay
processes are: flux tube (string) breaking, which favors large decay
widths for configurations with long flux tube components; and quantum
fluctuations that unlink flux tubes, which tends to broaden states
with short flux tube components. As yet we are not able to go beyond
providing a phenomenological fit to these qualitative
observations~\cite{Buniy:2002yx}, but hope to be able to do so in
the future.

In conclusion, let us return to continuum physics and consider a slab
of material that can support flux tubes. We have in mind a super-fluid
or superconductor, but are not limited to these possibilities. Assume
further that the flux tubes carry one and only one unit of flux. Next
consider manipulating these flux tubes. For instance, consider a
hypothetical superconductor where the flux tubes are pinned at the
bottom of the slab, say by being attracted to the poles of some
magnetic material, and at the top of the slab they are each associated
with the pole of a movable permanent magnet, perhaps a magnetic
whisker, or fine solenoid.  Assuming the tubes have time to relax to
tight configurations, the energy released should correspond to the
universal energy spectrum described above.  Another collection of
physical systems of potential interest are the atomic Bose-Einstein
condensates. For example, laser stirring of dilute ${ }^{87}$Rb atoms
at $80\ \textrm{nK}$ has produced vortices~\cite{BEC}, which
could lead to knots and links.

This work is supported in part by U.S. DoE grant \#~DE-FG05-85ER40226.

\end{document}